\documentclass{aastex}          
\usepackage{spr-astr-addons}    

\begin{document}
\def\gsim{\;\lower.6ex\hbox{$\sim$}\kern-6.7pt\raise.4ex\hbox{$>$}\;}
\def\lsim{\;\lower.6ex\hbox{$\sim$}\kern-6.7pt\raise.4ex\hbox{$<$}\;}
%
\title{Ultra Long Period Cepheids: \\
a primary standard candle out to the Hubble flow.}

\shorttitle{ULP Cepheids out to the Hubble flow}
\shortauthors{Fiorentino et al.}

\author{Fiorentino, G.\altaffilmark{1}}
\email{giuliana.fiorentino@oabo.inaf.it}
\and  
\author{Clementini, G.\altaffilmark{1}}
\and 
\author{Marconi, M.\altaffilmark{2}} 
\and 
\author{Musella, I\altaffilmark{2}}
\and 
\author{Saha, A.\altaffilmark{3}}
\and  
\author{Tosi, M.\altaffilmark{1}}
\and 
\author{Contreras Ramos, R.\altaffilmark{1}}
\and  
\author{Annibali, F.\altaffilmark{1}}
\and  
\author{Aloisi, A.\altaffilmark{4}}
\and  
\author{van der Marel, R.\altaffilmark{4}} 

\altaffiltext{1}{INAF- Osservatorio Astronomico di Bologna, via Ranzani 1, 40127, Bologna, Italy.}
\altaffiltext{2}{INAF- Osservatorio Astronomico di Capodimonte, vicolo
  Moiariello 16, 40127, Napoli, Italy.}
\altaffiltext{3}{National Optical Astronomy Observatory, P.O. Box 26732, Tucson, AZ 85726, USA}
\altaffiltext{4}{Space Telescope Science Institute, 3700 San Martin Drive, Baltimore, MD 21218, USA}

\begin{abstract}
The cosmological distance ladder crucially depends on classical
Cepheids (with P=3-80~days), which are primary distance indicators
up to 33~Mpc. Within this volume, very few SNe~Ia have been calibrated
through classical Cepheids, with uncertainty related to the non-linearity and the
metallicity dependence of their period$-$luminosity (PL) relation. Although a general consensus on these effects is still not achieved,
classical Cepheids remain the most used primary distance indicators. A possible
extension of these standard candles to further distances would be important. In this context, a very
promising new tool is represented by the ultra-long period (ULP) Cepheids
(P$\gsim$80~days), recently identified in star-forming
galaxies. Only a small number of ULP Cepheids have been discovered so far. Here we present and analyse the properties of an updated
sample of 37 ULP Cepheids observed in galaxies within a very large metallicity
range of 12$+$log(O/H) from $\sim$7.2 to 9.2 dex.
We find that their location in the colour(V$-$I)$-$magnitude diagram as well
as their Wesenheit (V$-$I) index-period (WP) relation suggests that
they are the counterparts at high luminosity of the shorter$-$period
(P $\lsim$ 80~days) classical Cepheids. However, a complete
pulsation and evolutionary theoretical scenario is needed to properly
interpret the true nature of these objects. 
We do not confirm the
flattening in the studied WP relation suggested by \citet{bird09}. Using the whole sample, we find that
 ULP Cepheids lie around a relation similar to that of the LMC, although with a large spread ($\sim$ 0.4 mag).
\end{abstract}


\section{Introduction}
The two most popular routes to estimating the Hubble costant
H$_0$ involve
the Cosmic Microwave Background (CMB) and the Supernovae type Ia
(SNe~Ia). The WMAP experiment \citep{komatsu11} measures a precise time since
recombination, and makes a strong case for a flat 
Universe. However, this measurement of the local expansion rate relies
on the adopted cosmological model and on the priors (list of
cosmological parameters) adopted in dealing with the CMB map. The
SNe~Ia provide an independent estimate of H$_0$. The absolute calibration
of the SNe~Ia luminosity peak is currently anchored to the
Cepheid-based distances to a dozen nearby host galaxies, and for
this reason Cepheids are the cornerstone for the absolute calibration
of the extragalactic distance scale
\citep[e.g.][]{freedman01,saha01}. However, the universality of the
Cepheid PL relation, and the possibility that the slope
and/or the zero-point of the PL relation might depend on the chemical
composition, have been lively debated for almost two decades
\citep[see e.g.][]{kennicutt98,fiorentino02,sakai04,marconi05,bono08,romaniello08} with
controversial results. No general consensus has been reached so
far. The consequence of this uncertainty is that the
most recent estimates of H$_0$, in spite of their very small formal
accuracy ($\sim$ 3\%), might
still be affected by systematic uncertainties of the order of 10\%. To prove this statement, we emphasize that the values presented in
this volume range from  62 to 74 Km/s Mpc. The above scenario can be
summarized quoting \citet{sandage09}: ``... the universality of the
PL relation is an only historically justified illusion.''

In this paper, we will discuss an on-going project devoted to the
  understanding of the nature of ULP Cepheids and their usefullness as
  stellar candles. In the next future, this possible new class of
  distance indicators will allow us to directly reach distances of
  cosmological interest, representing both an important alternative to
  SNe Ia and a better calibrator than classical Cepheids of this
  important secondary distance indicator as well as of other
  indicators. The reason of our interest in these stars is their intrinsic
luminosity, which is 2-4 magnitudes brighter than typical classical Cepheids used so far to set the extragalactic distance
ladder. Then, they allow us to extend the current observational limit
\citep[NGC~1309 is the farthest galaxy with identified classical
cepheids and it is located at 33~Mpc, see][]{riess09} up to 100~Mpc. Furthermore, the ESA astrometric satellite Gaia will
  provide trigonometric parallaxes at micro-arcsec accuracy, hence
  precise direct distances of the LMC and SMC ULP Cepheids. And last,
  but not least, the ULP's potential is to be further enhanced by new generation ground-based telescopes such as the
European Extremely Large Telescope \citep[E$-$ELT,][]{tolstoy10,deep11,fiorentino11}, which can make the ULP Cepheids (observable up to 320~Mpc) the first primary distance
indicator capable to directly measure H$_0$  in ``one$-$step''.

\section{The Ultra Long Period sample}\label{sec1}

\begin{figure}[t]
\includegraphics[width=\columnwidth]{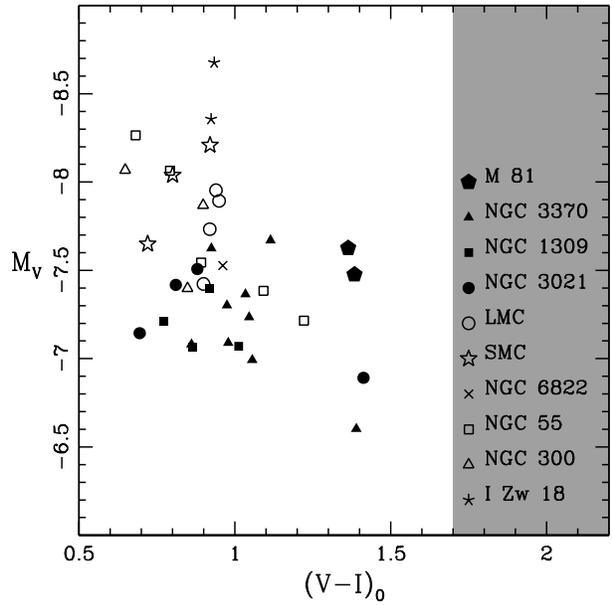}
\caption{%
Colour$-$magnitude diagram for the ULP Cepheids observed so far for which V
  and I photometry is available in the literature. Distance modulus and
  reddening used to plot ULP Cepheids in this plane are reported in
  Table~\ref{tab1}. Metal$-$rich galaxies (12+log(O/H) $\gsim$ 8.4
  dex or Z=0.008, i.e. LMC) have been
  highlighted with filled symbols.}
\label{fig1} 
\end{figure}

 \begin{table*}[t]
\centering
\scriptsize
 \caption{Compilation of ULP Cepheids observed in galaxies with a large range
   of metallicity. This table does not include 27 cepheids observed in
   NGC~4536 (4), NGC~4639 (2), NGC~5584 (7), NGC~4038 (8) and NGC~4258
   (6) with period longer than 80~days for which the V and I photometry
is not yet available \citep{riess11a}.} 
  \label{tab1}
 \begin{tabular}{lccccccc}
\tableline  
{Galaxy} & Period & V & V$-$I & $\mu_0$ & E(B-V) &Z & 12+log(O/H)\\
\hline
  & day & mag & mag & mag & mag & & dex \\
\hline
\multicolumn{8}{c}{18 ULP Cepheids compiled in Bird et al. 2009}\\
\hline
I~Zw~18 & 130.3 & 23.96 & 0.96 & 31.30 & 0.03 &  $\sim$0.0004 &7.21$^1$ \\
I~Zw~18 & 125.0 & 23.65 & 0.97 & 31.30 & 0.03 &  $\sim$0.0004 & 7.21$^1$ \\
SMC & 210.4 & 12.28 & 0.83 & 18.93 & 0.09 & $\sim$0.002 & 7.98$^2$ \\
SMC & 127.5 & 11.92 & 1.03 & 18.93 & 0.09 & $\sim$0.002 &  7.98$^2$ \\
SMC & 84.4 & 11.97 & 0.91 & 18.93 & 0.09 & $\sim$0.002 &  7.98$^2$ \\
NGC~55 & 175.9 & 19.25 & 0.84 & 26.43 & 0.13 & $\sim$0.003 &  8.05$^3$ \\
NGC~55 & 152.1 & 19.56 & 0.95 & 26.43 & 0.13 & $\sim$0.003 & 8.05$^3$ \\
NGC~55 & 112.7 & 20.18 & 1.05 & 26.43 & 0.13 & $\sim$0.003 & 8.05$^3$ \\
NGC~55 & 97.7 & 20.54 & 1.25 & 26.43 & 0.13 &  $\sim$0.003 &8.05$^3$ \\
NGC~55 & 85.1 & 20.84 & 1.38 & 26.43 & 0.13 & $\sim$0.003 & 8.05$^3$ \\
NGC~6822 & 123.9 & 17.86 & 1.40 & 23.31 & 0.36 & $\sim$0.003& 8.11$^4$ \\
NGC~300 & 115.8 & 20.13 & 0.97 & 26.37 & 0.10 & $\sim$0.004 & 8.25$^5$ \\
NGC~300 & 89.1 & 19.71 & 1.02 & 26.37 & 0.10 & $\sim$0.004 & 8.25$^5$ \\
NGC~300 & 83.0 & 19.26 & 0.77 & 26.37 & 0.10 & $\sim$0.004 &8.25$^5$\\ 
LMC & 118.7 & 11.99 & 1.12 & 18.50 & 0.14 & $\sim$0.008 & 8.39$^6$ \\
LMC & 109.2 & 12.41 & 1.07 & 18.50 & 0.14 & $\sim$0.008 & 8.39$^6$ \\
LMC & 98.6 & 11.92 & 1.11 & 18.50 & 0.14 & $\sim$0.008 & 8.39$^6$ \\
LMC & 133.6 & 12.12 & 1.09 & 18.50 & 0.14 & $\sim$0.008 & 8.39$^6$ \\
\hline
\multicolumn{8}{c}{19 ULP Cepheids compiled in this paper}\\
\hline
M~81 & 96.766 & 21.52 & 1.40 & 27.69 & 0.08  & $\sim$0.013 & 8.7$^7$      \\
M~81 & 98.981 & 21.69 & 1.42 & 27.69 & 0.08  & $\sim$0.013 & 8.7$^7$      \\
NGC~3370 & 80.85 & 26.002 & 0.90 & 32.13 & 0.03 &  $\sim$0.02 & 8.82 (9.21)$^8$  \\
NGC~3370 & 81.04 & 25.895 & 1.01 & 32.13 & 0.03 &  $\sim$0.02 & 8.82 (8.92)$^8$ \\
NGC~3370 & 83.28 & 27.010 & 1.43 & 32.13 & 0.03 &  $\sim$0.02 & 8.82 (8.66)$^8$ \\
NGC~3370 & 86.33 & 25.892 & 1.07 & 32.13 & 0.03 &  $\sim$0.02 & 8.82 (8.93)$^8$ \\
NGC~3370 & 88.25 & 26.112 & 1.01 & 32.13 & 0.03 &  $\sim$0.02 & 8.82 (8.70)$^8$ \\
NGC~3370 & 88.54 & 25.667 & 1.15 & 32.13 & 0.03 &  $\sim$0.02 & 8.82 (9.24)$^8$ \\
NGC~3370 & 96.49 & 25.522 & 0.96 & 32.13 & 0.03 &  $\sim$0.02 & 8.82 (8.89)$^8$ \\
NGC~3370 & 96.82 & 26.286 & 1.09 & 32.13 & 0.03 &  $\sim$0.02 & 8.82 (8.66)$^8$ \\
NGC~3370 & 98.72 & 26.034 & 1.08 & 32.13 & 0.03 &  $\sim$0.02 & 8.82 (9.04)$^8$ \\
NGC~1309 & 82.13 & 26.481 & 0.90 & 32.59 & 0.04 &  $\sim$0.02 & 8.90 (8.87)$^8$ \\
NGC~1309 & 82.38 & 26.246 & 0.81 & 32.59 & 0.04 &  $\sim$0.02 & 8.90 (9.07)$^8$ \\
NGC~1309 & 89.03 & 26.625 & 1.05 & 32.59 & 0.04 &  $\sim$0.02 & 8.90 (9.25)$^8$ \\
NGC~1309 & 97.89 & 26.201 & 0.95 & 32.59 & 0.04  &  $\sim$0.02 & 8.90 (9.13)$^8$ \\
NGC~3021 & 82.66 & 25.913 & 0.73 & 32.27 & 0.013  &  $\sim$0.02 & 8.94 (8.65)$^8$\\
NGC~3021 & 88.18 & 26.884 & 1.45 & 32.27 & 0.013  &  $\sim$0.02 & 8.94 (9.10)$^8$\\
NGC~3021 & 90.73 & 25.735 & 0.92 & 32.27 & 0.013  &  $\sim$0.02 & 8.94 (9.14)$^8$\\
NGC~3021 & 95.91 & 25.756 & 0.85 & 32.27 & 0.013  &  $\sim$0.02 & 8.94 (8.94)$^8$\\
\hline
\end{tabular}
%
\tablenotetext{1}{\citet{fiorentino10a}, \citet{contreras11};
$^2$ \citet{peimbert76}, \citet{hilditch05}, \citet{keller06}; $^3$
\citet{pietrzynski06}; $^4$ \citet{pietrzynski04}; $^5$
\citet{gieren04}, \citet{gieren05}, \citet{urbaneja05}; $^6$
\citet{freedman85}, \citet{pagel78}, \citet{udalski99}; $^7$
\citet{gerke11}; $^8$\citet{riess09}.}
\end{table*}

The recent identification of ULP Cepheids can be attributed to different observational biases
conspiring to make them elusive for such a long time.  
In fact, they may have often escaped detection because of:
\begin{itemize} 
\item the very long time baseline (up to few years)
needed to well characterize their periods; 
\item their very bright intrinsic luminosity that causes them to be
saturated sources in the photometry of closeby galaxies. 
\end{itemize}
This is the case in the Magellanic Clouds whose ULP Cepheids are in fact very
close to the saturation limit of the OGLE survey \citep{soszynski08b}. 
\hspace{-2cm}
%


\citet{bird09} collected 18 ULP Cepheids, with period longer than 80~days, in six observed
metal$-$poor (12+log(O/H)$\lsim$8.4 dex or Z=0.008\footnote{
    Assuming
that the oxygen abundance is a very robust proxy of the iron abundance
(i.e., [Fe/H] = [O/H]), we can use the following relation to convert
12+log(O/H) in Z:
Z=Z$_{\odot}\times$10$^{(log(O/H)-12+3.1)}$. To
compute the values in Table~\ref{tab1}, we have assumed log(O/H)$_{\odot}$ = -3.10, Z$_{\odot}$=0.02 and Y$_{\odot}$=0.27.}) star forming galaxies, namely the Magellanic Clouds, NGC~6822,
NGC~55, NGC~300 and I~Zw~18 (see Table~\ref{tab1}). Using available V, I
photometry they studied the ULP Cepheid location in the logP vs V and
I filters and the logP vs the V$-$I Wesenheit
  index\footnote{ The Wesenheit index is defined 
as $W = I - 1.55\times(V-I)$, where $V$ and $I$ are the apparent
magnitudes.} index diagrams in comparison with classical Cepheids in the Small
Magellanic Cloud (SMC). The authors noticed that this sample shows a
large scatter around the relationships that hold for SMC classical
Cepheids. They suggested that this scatter could be due to a possible
flattening of the PL relations and in particular of the WP
relations and they suggest that ULP Cepheids could be better standard
candles than classical Cepheids.
More recently, other 46 ULP Cepheids have been detected in predominantly
metal$-$rich (12+log(O/H)$\gsim$8.4 dex or Z=0.008) galaxies where SNe~Ia were observed. For 19 ULP Cepheids out of
46, V and I photometry is available in literature. 

We considered the properties of the updated sample of ULP Cepheids (see Table~\ref{tab1}). 
It is worth mentioning that the ULP sample is going to further
increase in the next years thanks to the opportunity to follow up
the surveyed galaxies to minimize the errors on the calibration of SNe
Ia using
classical Cepheids \citep[][and references therein]{riess11a}.
In fact, in this context 27 new objects have been observed and
classified as ULP Cepheids with Wide Field Camera 3 on board HST. However, in
the literature,
only F160W photometry and V$-$I colour are available. Furthermore, we observed the Blue Compact Galaxy NGC~1705 with Gemini$-$South, for which a detailed star formation history is known throught a careful analysis of
UBVIJH data taken with WFPC2, ACS and NICMOS on board of HST \citep{tosi01,annibali03,annibali09}. In this galaxy, by analogy with I~Zw~18, we expect 
to observe ULP Cepheids (in addition to classical Cepheids), given the occurence of several stars with mass
around and above 15 M$_{\odot}$ and thus to have the opportunity to constrain the
ULP Cepheids evolutionary state (see Sect.~\ref{sec3}), these results
will be soon published (Fiorentino et al. in prep.). We have also obtained time on the Gemini$-$North telescope and on the Telescopio
  Nazionale Galileo to follow up the most metal$-$poor ULP Cepheids found so far in I~Zw~18
\citep{fiorentino10a} and a long period cepheid observed in UGC~9128 \citep[or DDO~187][]{hoessel98}.  
These two galaxies are the only ones containing ULP Cepheids
that have a metallicity lower than the
typical value for the SMC (12+log(O/H)$\lsim$ 8 dex or Z=0.004). On this basis they are
crucial to constraints
the evolutionary and pulsation models of ULP Cepheids (see Sect.~\ref{sec3}).

In Fig.~\ref{fig1}, the location in the colour$-$magnitude
diagram of the whole sample collected in this paper is shown. In order to emphasize the metallicity dependence
of the Cepheid location
 we have used empty and starred symbols for metal$-$poor 
galaxies and filled ones for metal$-$rich galaxies.
With only a few exceptions and in spite of the large
range of metallicity, most of the ULP Cepheids seem to
occupy the same region of the instability strip (IS). Furthermore, the main difference between ULP Cepheids belonging to different
galaxies seems to be the intrinsic luminosity, the metal$-$rich
ULP Cepheids also being the fainter ones. These metal$-$rich ULP
Cepheids also show shorter periods (P $\lsim$ 100 days) than the
metal$-$poor ones.
In order to properly describe this behaviour, we are developping a full
pulsation and evolutionary theoretical
scenario for these very bright pulsators with metallicities ranging
from Z=0.0004 to Z=0.02 taking advantage of our experience in modeling
classical Cepheids \citep{bono97a,fiorentino02,marconi05}.

\section{Period {\it vs} the Wesenheit index}\label{sec2}

\begin{figure}[t]
\includegraphics[width=\columnwidth]{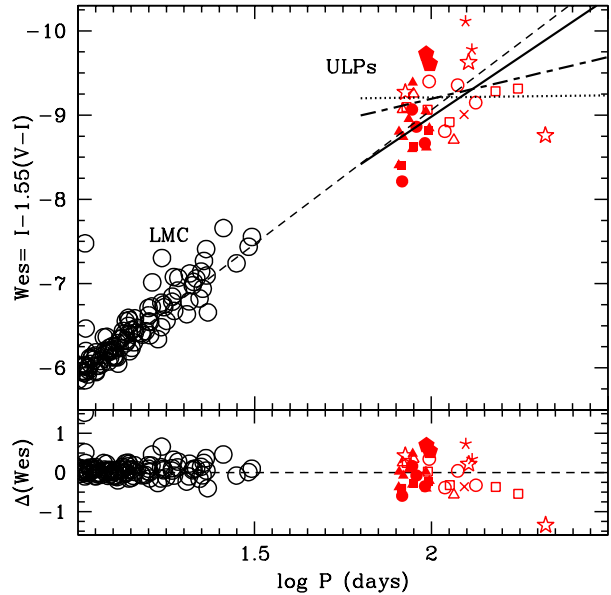}
\caption{%
{\it Top-panel:} The logP vs Wesenheit index in V and I bands. Black dots are the LMC classical Cepheids
whereas red empty and filled dots are metal$-$poor and metal$-$rich
ULP Cepheids recently observed (see Table~\ref{tab1}, for details). Dashed line
represents the classical Cepheid WP relation for the LMC
extrapolated to long period range. Dotted line
represent the flat slope suggested by \citet{bird09} using only
the metal$-$poor sample. Dotted--dashed line represent the same linear fit excluding the very long period ULP Cepheid in the SMC (star). Solid line shows the fit
obtained adding the whole sample collected in this paper. {\it Bottom-panel:} The residual
to the LMC relation have been shown to highlight the
significant spread at high luminosity.  {\it Both-panels}: the symbol$-$code for ULP
Cepheids is the same
used in Fig.~\ref{fig1}.}
\label{fig2} 
\end{figure}

The Cepheid distances are usually derived by means of the WP relation \citep{madore76,fiorentino07,bono08}. This formulation of the
more general PL relation has the advantage to be reddening$-$free by
construction and to include the colour  (and in turn the temperature)
information for each individual star, defining a
very tight relation. In their work, \citet{bird09} discussed the WP relation of the ULP sample and compared it to that of
classical Cepheids in the SMC. They found that ULP Cepheids have a larger
scatter ($\sim$0.5 mag) 
when compared with SMC classical Cepheids ($\lsim$ 0.08 mag).    
They claim that this large scatter may depend on the poor ULP's 
sample used in their paper, suggesting that more observations are
needed in order to collect a significant statistical sample that could compete
with the accuracy reached by the OGLE III survey for the SMC
\citep[which includes 2626 fundamental mode
Cepheids,][]{soszynski10a}. \citet{bird09} also suggest that a
metallicity dependence could exist and cause the large scatter observed in the WP relation, as the investigated variables cover a  metallicity
range from 12 + log(O/H) = 7.22 to
8.39 dex. However, they conclude that the sample is affected by distance
uncertainty and that the metallicity effect can not be proven. 
Moreover, most of the variables used by \citet{bird09} are in galaxies with 
metallicities similar or poorer than SMC (Z=0.004, 12 + log(O/H)= 7.98
dex), so in a metallicity range where this dependence is expected to
be small. The only exception
is represented by the two ULP Cepheids recently discovered in the extremely
metal-poor (Z=0.0004, 12 + log(O/H)=7.21 dex) Blue Compact Dwarf galaxy
I~Zw~18 \citep[][see Sect.~\ref{sec3} for details]{fiorentino10a}. They also notice that I~Zw~18 Cepheids have the
largest scatter in the PL plane in both V and I bands and, for this
reason, they do not
use them in their analysis. Concerning the metallicity effect on the WP relation, we note
here that \citet{bono10b} using updated theoretical models for classical Cepheids
\citep{bono99,fiorentino02,marconi05} show that the WP relation
in V and I bands is not supposed to be metal$-$dependent and that
Marconi et al. (2010) predict an insensitivity to metallicity 
when the metallicity decreases below the typical SMC  value. However, a recent application of the infrared surface brightness
technique to Cepheids of SMC, LMC and the Galaxy shows an opposite
effect \citep{storm11}. 

\begin{figure}[t]
\includegraphics[width=\columnwidth]{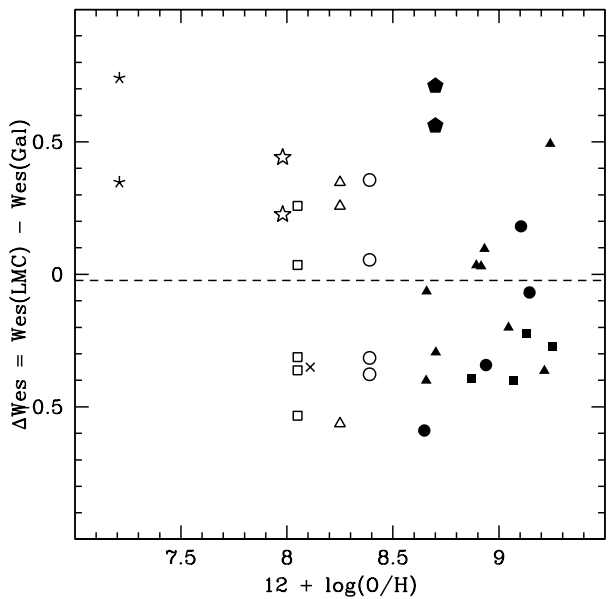}
\caption{%
Metallicity$-$dependence of the Wesenheit index. The difference between
the Wesenheit index, as defined in the LMC, and in the host galaxy is
plotted against the
metallicity [12 + log(O/H)]. The symbol$-$code is the same
used in Fig.~\ref{fig1}.}
\label{fig3} 
\end{figure}

In Fig.~\ref{fig2}, we show the WP relation for the updated ULP sample in comparison
with the LMC classical Cepheids, released from the
OGLE III survey \citep[$\sim$~1849 stars pulsating in the fundamental mode,][]{soszynski08b}. Using only galaxies with 12 +
log(O/H) $\le$ 8.4 dex, Bird and collaborators found an almost flat
slope, i.e. Wes(I, V$-$I)=$-$9.12$-$0.05 log P
with $\sigma$=0.36 mag (dotted line in Fig~\ref{fig2}). However, excluding from the fit the SMC cepheid
with the longest period, we find Wes(I, V$-$I)=$-$7.21$-$0.99 log P with $\sigma$=0.34 mag (dashed$-$dotted line in
Fig.~\ref{fig2}).  If we include in the fit
the ULP Cepheids in the metal$-$richer galaxies, approaching the solar
metallicity such as
NGC~1309, NGC~3370, NGC~3021 and M~81 \citep[recently observed
by][]{riess09,gerke11}, we see that a slope is drawn corresponding to 
Wes(I, V$-$I)=$-$3.68$-$2.66 log P with $\sigma =$0.34 mag (solid line),
which starts to resamble the Wesenheit index observed in the LMC,
i.e. Wes(I, V$-$I)= $-$2.70$-$3.187 log P with $\sigma =$0.08 mag (dashed line).

Our updated sample covers a larger metallicity
range, with 12$+$log(O/H) going from 7.21 dex to 8.94 dex. Thus, we are able
to extend the work presented in \citet{bird09} and to better investigate the possible
metallicity effect on the Wesenheit index. We have computed the
difference between the Wesenheit index
expected for the LMC and the observed one. In Fig.~\ref{fig3} we show this difference versus the
metallicity for each galaxy. For metal$-$rich ULP Cepheids, we have taken into
account the metallicity gradient measured in the host
galaxies \citep{riess09} instead of using the mean metallicities. In order to do this we have estimated the distance of each ULP
from the centre of the galaxy and then we have assigned to the star a
metallicity value derived using the formula presented in \citet{riess09}. In Table~\ref{tab1}, we reported both the mean and the
individual metallicity (in parenthesis) for each star. The derived
metallicities span a larger range that goes from 12$+$log(O/H)=8.6 to
9.2 dex.

From Fig.~\ref{fig3} we can only draw some qualitative
conclusions. In fact, we are aware that the ``indirect'' metallicity
proxy [O/H] is not the best measurement we can use to  identify the
metallicity effect,  and that individual spectroscopic measurements
\citep{romaniello08,pedicelli09} would be necessary to this purpose. 

No significant trend with metallicity is observed. Indeed all the
samples cluster around an almost zero average (dashed
line in Fig.~\ref{fig3}) apart from the ones of I~Zw~18 and M~81 that could be affected by blending with blue and red stars
respectively \citep{fiorentino10a,gerke11}.

\section{Pulsation and evolutionary models in the low metallicity regime}\label{sec3}

\begin{figure}[t]
\includegraphics[width=\columnwidth]{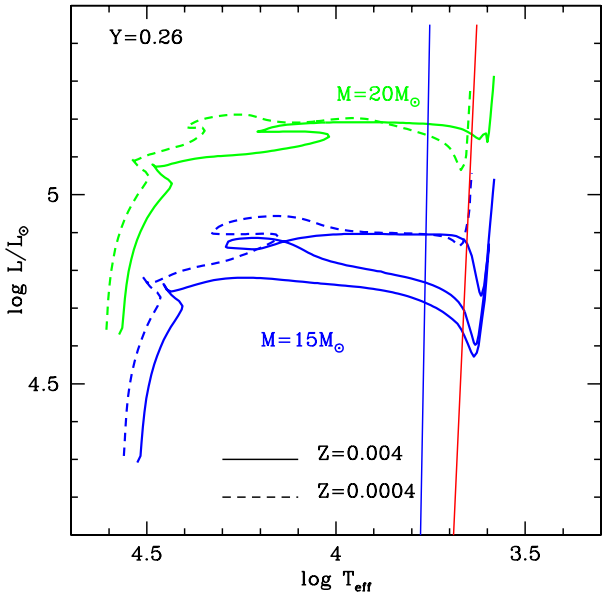}
\caption{%
Evolutionary tracks for 15 and 20
M$_{\odot}$ from the Padua database \citep{bertelli09} plotted in the
Hertzsprung$-$Russel diagram. Solid lines represents a metallicity
typical for SMC whereas dashed line
represent the very low metallicity typical of I~Zw~18. The two almost vertical solid lines represent the
instability strip as computed using updated theoretical models ``ad
hoc" for I~Zw~18 \citep{marconi10}.}
\label{fig4} 
\end{figure}

The discovery of two bona-fide ULP Cepheids with periods of 125 and 130.3
days in I~Zw~18 poses some problems on the theoretical interpretation
of these objects. The ULP Cepheids lie in a region of the HR diagram where
current stellar evolution models do not predict the existence of
core-helium burning blue loops, the evolutionary phase which so far produces Cepheids. 

As an example, in Fig.~4 we show the evolutionary tracks from \citet{bertelli09}
for M=15 and 20 M$_{\odot}$ in the low metallicity regime, namely for
Z=0.004 (or 12+log(O/H)$\sim$ 8 dex, solid line) and Z=0.0004 (or
12+log(O/H)$\sim$ 7.2 dex, dashed line). These masses bracket the
luminosity observed for the ULP Cepheids in I~Zw~18. We have also reported the red
(cold) and blue (warm) theoretical boundaries of the pulsation
IS as presented in \citet{marconi10}. 
It is clear that the only evolutionary track that shows a ``classical''
blue loop crossing the IS is the track for 15 M$_{\odot}$ and Z=0.004.
The remaining evolutionary tracks cross the IS only after, and not during, the central helium burning phase. This has
some effect on the evolutionary times spent in the IS
and in turn on the probability that we have to observe these objects. Here
it is worth to mention that the blue loop phase 
has to be treated with caution. In fact, its occurrence is related to
different physical parameters. The most important are:
{\it i)} the chemical composition which influences the extension of
the blue loop. The loop is more extended when the helium content increases (at fixed
metallicity) whereas it is less extended when
the metal content increases (at fixed helium content); {\it ii)} the
mixing during the Hydrogen central burning that reduces the
extension of the blue loop if its efficiency increases. 

In this context the observation of pulsating objects can give sound
constraints on the involved physical mechanisms, in particular when
connected with the star formation history of the host galaxy. In fact,
star counts are directly correlated to the evolutionary times spent in the 
different phases. The computations of new evolutionary models is in
progress in order to provide firm constraints both on the evolutionary
times and on the inputs for our non-linear convective pulsation
models. We note that the full understanding of the theoretical
scenario in the metal-poor regime is particularly important. In fact,
at these low metallicities (Z$\lsim$ 0.008), theoretical models
predict the long periods observed
for ULP Cepheids only assuming stellar masses and luminosities
significantly higher than that characterizing classical
Cepheids. Therefore, a pure extrapolation of classical Cepheids properties
such as the mass-luminosity relation and the
inner structure seems not to hold for metal$-$poor ULP Cepheids pulsation
models \citep[see][]{marconi10}.

\section{Conclusions}

We have updated the sample of ULP Cepheids available in literature
extending the covered metallicity range which now spans $\sim$~2 dex, from
12$+$log(O/H)$=$7.23 dex to 9.2 dex. The main results are
the following:
\begin{itemize}
\item The observed properties, such as the location in the
colour$-$magnitude diagram and the WP relation, suggest that they are the counterparts at high
luminosity of the shorter$-$period (P $\lsim$ 80~days) classical
Cepheids. 
\item Using the whole updated sample, we do not confirm the flattening in the Wesehneit index suggested by \citet{bird09}. Instead we find that:\\
Wes(I, V$-$I)=$-$3.68$-$2.66 log P\\
with $\sigma =$0.34 mag. Then, these pulsators lie with a large spread ($\sim$ 0.4
mag) around a WP relation consistent within the errors with that of the LMC Cepheids. 
\item We have qualitatively investigated the metallicity dependence of the Wesenheit index derived from the V and I bands. We do not find a significant metallicity dependence in good agreement with the theoretical modelling of
classical Cepheids \citep{bono10b}. Thus, the metallicity
differences do not seem to be the cause of the spread around the WP relationship.
The cause could likely be the non$-$homogeneity and the poor statistic of the
sample, but also mean magnitudes and colours not well determined. 
\end{itemize}
Concluding, we suggest to follow up galaxies where ULP Cepheids have
been observed and to perform variability studies
in particular in the metal$-$poor range (12 + log(O/H) $\lsim$ 8.4
dex). A new homogeneous and larger sample will help us to understand
the nature of these objects. In fact the
observation of such massive objects (from 15 to 20 M$_{\odot}$) in
the low metallicity regime poses some problems in their interpretation
as blue$-$loop stars crossing the classical IS and their existence gives also some constrain to the internal structure of these objects.
On the other hand the construction of new accurate evolutionary and
pulsation models for ranges of masses corresponding to very metal poor
ULP Cepheids will allow us to understand the true status as these pulsators and
to confirm their nature of cosmological standard candles.

%

%
%

%

%
%

%

%
\acknowledgments

This work has been entirely supported by the INAF fellowship 2009 grant.


%
\bibliographystyle{spr-mp-nameyear-cnd}  

\begin{thebibliography}{0}
\ifx \bisbn   \undefined \def \bisbn  #1{ISBN #1}\fi
\ifx \binits  \undefined \def \binits#1{#1} \fi
\ifx \bauthor  \undefined \def \bauthor#1{#1} \fi
\ifx \batitle  \undefined \def \batitle#1{#1} \fi
\ifx \bjtitle  \undefined \def \bjtitle#1{#1}\fi
\ifx \bvolume  \undefined \def \bvolume#1{\textbf{#1}}\fi
\ifx \byear  \undefined \def \byear#1{#1} \fi
\ifx \bissue  \undefined \def \bissue#1{#1} \fi
\ifx \bfpage  \undefined \def \bfpage#1{#1} \fi
\ifx \blpage  \undefined \def \blpage #1{#1} \fi
\ifx \burl  \undefined \def \burl#1{\textsf{#1}} \fi
\ifx \doiurl  \undefined \def \doiurl#1{\textsf{#1}} \fi
\ifx \betal  \undefined \def \betal{\textit{et al.}} \fi
\ifx \binstitute  \undefined \def \binstitute#1{#1} \fi
\ifx \binstitutionaled  \undefined \def \binstitutionaled#1{#1} \fi
\ifx \bctitle  \undefined \def \bctitle#1{#1} \fi
\ifx \beditor  \undefined \def \beditor#1{#1} \fi
\ifx \bpublisher  \undefined \def \bpublisher#1{#1} \fi
\ifx \bbtitle  \undefined \def \bbtitle#1{#1} \fi
\ifx \bedition  \undefined \def \bedition#1{#1} \fi
\ifx \bseriesno  \undefined \def \bseriesno#1{#1} \fi
\ifx \blocation  \undefined \def \blocation#1{#1} \fi
\ifx \bsertitle  \undefined \def \bsertitle#1{#1} \fi
\ifx \bsnm \undefined \def \bsnm#1{#1} \fi
\ifx \bsuffix \undefined \def \bsuffix#1{#1} \fi
\ifx \bparticle \undefined \def \bparticle#1{#1} \fi
\ifx \barticle \undefined \def \barticle#1{#1} \fi
\ifx \bconfdate \undefined \def \bconfdate #1{#1} \fi
\ifx \botherref \undefined \def \botherref #1{#1} \fi
\ifx \url \undefined \def \url#1{\textsf{#1}} \fi
\ifx \bchapter \undefined \def \bchapter#1{#1} \fi
\ifx \bbook \undefined \def \bbook#1{#1} \fi
\ifx \bcomment \undefined \def \bcomment#1{#1} \fi
\ifx \oauthor \undefined \def \oauthor#1{#1} \fi
\ifx \citeauthoryear \undefined \def \citeauthoryear#1{#1} \fi
\ifx \endbibitem  \undefined \def \endbibitem {}\fi
\ifx \bconflocation  \undefined \def \bconflocation#1{#1} \fi
\ifx \arxivurl  \undefined \def \arxivurl#1{\textsf{#1}} \fi

\end{thebibliography}


\begin{thebibliography}{45}
\ifx \bisbn   \undefined \def \bisbn  #1{ISBN #1}\fi
\ifx \binits  \undefined \def \binits#1{#1} \fi
\ifx \bauthor  \undefined \def \bauthor#1{#1} \fi
\ifx \batitle  \undefined \def \batitle#1{#1} \fi
\ifx \bjtitle  \undefined \def \bjtitle#1{#1}\fi
\ifx \bvolume  \undefined \def \bvolume#1{\textbf{#1}}\fi
\ifx \byear  \undefined \def \byear#1{#1} \fi
\ifx \bissue  \undefined \def \bissue#1{#1} \fi
\ifx \bfpage  \undefined \def \bfpage#1{#1} \fi
\ifx \blpage  \undefined \def \blpage #1{#1} \fi
\ifx \burl  \undefined \def \burl#1{\textsf{#1}} \fi
\ifx \doiurl  \undefined \def \doiurl#1{\textsf{#1}} \fi
\ifx \betal  \undefined \def \betal{\textit{et al.}} \fi
\ifx \binstitute  \undefined \def \binstitute#1{#1} \fi
\ifx \binstitutionaled  \undefined \def \binstitutionaled#1{#1} \fi
\ifx \bctitle  \undefined \def \bctitle#1{#1} \fi
\ifx \beditor  \undefined \def \beditor#1{#1} \fi
\ifx \bpublisher  \undefined \def \bpublisher#1{#1} \fi
\ifx \bbtitle  \undefined \def \bbtitle#1{#1} \fi
\ifx \bedition  \undefined \def \bedition#1{#1} \fi
\ifx \bseriesno  \undefined \def \bseriesno#1{#1} \fi
\ifx \blocation  \undefined \def \blocation#1{#1} \fi
\ifx \bsertitle  \undefined \def \bsertitle#1{#1} \fi
\ifx \bsnm \undefined \def \bsnm#1{#1} \fi
\ifx \bsuffix \undefined \def \bsuffix#1{#1} \fi
\ifx \bparticle \undefined \def \bparticle#1{#1} \fi
\ifx \barticle \undefined \def \barticle#1{#1} \fi
\ifx \bconfdate \undefined \def \bconfdate #1{#1} \fi
\ifx \botherref \undefined \def \botherref #1{#1} \fi
\ifx \url \undefined \def \url#1{\textsf{#1}} \fi
\ifx \bchapter \undefined \def \bchapter#1{#1} \fi
\ifx \bbook \undefined \def \bbook#1{#1} \fi
\ifx \bcomment \undefined \def \bcomment#1{#1} \fi
\ifx \oauthor \undefined \def \oauthor#1{#1} \fi
\ifx \citeauthoryear \undefined \def \citeauthoryear#1{#1} \fi
\ifx \endbibitem  \undefined \def \endbibitem {}\fi
\ifx \bconflocation  \undefined \def \bconflocation#1{#1} \fi
\ifx \arxivurl  \undefined \def \arxivurl#1{\textsf{#1}} \fi

\bibitem[\protect\citeauthoryear{{Annibali} et~al.}{2003}]{annibali03}
\begin{barticle}
\bauthor{\bsnm{{Annibali}}, \binits{F.}},
\bauthor{\bsnm{{Greggio}}, \binits{L.}},
\bauthor{\bsnm{{Tosi}}, \binits{M.}},
\bauthor{\bsnm{{Aloisi}}, \binits{A.}},
\bauthor{\bsnm{{Leitherer}}, \binits{C.}}:
\bjtitle{\aj}
\bvolume{126},
\bfpage{2752}
(\byear{2003}).
\arxivurl{arXiv:astro-ph/0309265}.
doi:\doiurl{10.1086/379556}
\end{barticle}
\endbibitem

\bibitem[\protect\citeauthoryear{{Annibali} et~al.}{2009}]{annibali09}
\begin{barticle}
\bauthor{\bsnm{{Annibali}}, \binits{F.}},
\bauthor{\bsnm{{Tosi}}, \binits{M.}},
\bauthor{\bsnm{{Monelli}}, \binits{M.}},
\bauthor{\bsnm{{Sirianni}}, \binits{M.}},
\bauthor{\bsnm{{Montegriffo}}, \binits{P.}},
\bauthor{\bsnm{{Aloisi}}, \binits{A.}},
\bauthor{\bsnm{{Greggio}}, \binits{L.}}:
\bjtitle{\aj}
\bvolume{138},
\bfpage{169}
(\byear{2009}).
\arxivurl{0904.3257}.
doi:\doiurl{10.1088/0004-6256/138/1/169}
\end{barticle}
\endbibitem

\bibitem[\protect\citeauthoryear{{Bertelli} et~al.}{2009}]{bertelli09}
\begin{barticle}
\bauthor{\bsnm{{Bertelli}}, \binits{G.}},
\bauthor{\bsnm{{Nasi}}, \binits{E.}},
\bauthor{\bsnm{{Girardi}}, \binits{L.}},
\bauthor{\bsnm{{Marigo}}, \binits{P.}}:
\bjtitle{\aap}
\bvolume{508},
\bfpage{355}
(\byear{2009}).
\arxivurl{0911.2419}.
doi:\doiurl{10.1051/0004-6361/200912093}
\end{barticle}
\endbibitem

\bibitem[\protect\citeauthoryear{{Bird} et~al.}{2009}]{bird09}
\begin{barticle}
\bauthor{\bsnm{{Bird}}, \binits{J.C.}},
\bauthor{\bsnm{{Stanek}}, \binits{K.Z.}},
\bauthor{\bsnm{{Prieto}}, \binits{J.L.}}:
\bjtitle{\apj}
\bvolume{695},
\bfpage{874}
(\byear{2009}).
\arxivurl{0807.4933}.
doi:\doiurl{10.1088/0004-637X/695/2/874}
\end{barticle}
\endbibitem

\bibitem[\protect\citeauthoryear{{Bono} et~al.}{1997}]{bono97a}
\begin{barticle}
\bauthor{\bsnm{{Bono}}, \binits{G.}},
\bauthor{\bsnm{{Caputo}}, \binits{F.}},
\bauthor{\bsnm{{Cassisi}}, \binits{S.}},
\bauthor{\bsnm{{Castellani}}, \binits{V.}},
\bauthor{\bsnm{{Marconi}}, \binits{M.}}:
\bjtitle{\apj}
\bvolume{479},
\bfpage{279}
(\byear{1997}).
\arxivurl{arXiv:astro-ph/9609153}.
doi:\doiurl{10.1086/303872}
\end{barticle}
\endbibitem

\bibitem[\protect\citeauthoryear{{Bono} et~al.}{1999}]{bono99}
\begin{barticle}
\bauthor{\bsnm{{Bono}}, \binits{G.}},
\bauthor{\bsnm{{Caputo}}, \binits{F.}},
\bauthor{\bsnm{{Castellani}}, \binits{V.}},
\bauthor{\bsnm{{Marconi}}, \binits{M.}}:
\bjtitle{\apj}
\bvolume{512},
\bfpage{711}
(\byear{1999}).
\arxivurl{arXiv:astro-ph/9809127}.
doi:\doiurl{10.1086/306815}
\end{barticle}
\endbibitem

\bibitem[\protect\citeauthoryear{{Bono} et~al.}{2008}]{bono08}
\begin{barticle}
\bauthor{\bsnm{{Bono}}, \binits{G.}},
\bauthor{\bsnm{{Caputo}}, \binits{F.}},
\bauthor{\bsnm{{Fiorentino}}, \binits{G.}},
\bauthor{\bsnm{{Marconi}}, \binits{M.}},
\bauthor{\bsnm{{Musella}}, \binits{I.}}:
\bjtitle{\apj}
\bvolume{684},
\bfpage{102}
(\byear{2008}).
\arxivurl{0805.1592}.
doi:\doiurl{10.1086/589965}
\end{barticle}
\endbibitem

\bibitem[\protect\citeauthoryear{{Bono} et~al.}{2010}]{bono10b}
\begin{barticle}
\bauthor{\bsnm{{Bono}}, \binits{G.}},
\bauthor{\bsnm{{Caputo}}, \binits{F.}},
\bauthor{\bsnm{{Marconi}}, \binits{M.}},
\bauthor{\bsnm{{Musella}}, \binits{I.}}:
\bjtitle{\apj}
\bvolume{715},
\bfpage{277}
(\byear{2010}).
\arxivurl{1004.0363}.
doi:\doiurl{10.1088/0004-637X/715/1/277}
\end{barticle}
\endbibitem

\bibitem[\protect\citeauthoryear{{Contreras Ramos} et~al.}{2011}]{contreras11}
\begin{barticle}
\bauthor{\bsnm{{Contreras Ramos}}, \binits{R.}},
\bauthor{\bsnm{{Annibali}}, \binits{F.}},
\bauthor{\bsnm{{Fiorentino}}, \binits{G.}},
\bauthor{\bsnm{{Tosi}}, \binits{M.}},
\bauthor{\bsnm{{Aloisi}}, \binits{A.}},
\bauthor{\bsnm{{Clementini}}, \binits{G.}},
\bauthor{\bsnm{{Marconi}}, \binits{M.}},
\bauthor{\bsnm{{Musella}}, \binits{I.}},
\bauthor{\bsnm{{Saha}}, \binits{A.}},
\bauthor{\bsnm{{van der Marel}}, \binits{R.P.}}:
\bjtitle{\apj}
\bvolume{739},
\bfpage{74}
(\byear{2011}).
\arxivurl{1106.5613}.
doi:\doiurl{10.1088/0004-637X/739/2/74}
\end{barticle}
\endbibitem

\bibitem[\protect\citeauthoryear{{Deep} et~al.}{2011}]{deep11}
\begin{barticle}
\bauthor{\bsnm{{Deep}}, \binits{A.}},
\bauthor{\bsnm{{Fiorentino}}, \binits{G.}},
\bauthor{\bsnm{{Tolstoy}}, \binits{E.}},
\bauthor{\bsnm{{Diolaiti}}, \binits{E.}},
\bauthor{\bsnm{{Bellazzini}}, \binits{M.}},
\bauthor{\bsnm{{Ciliegi}}, \binits{P.}},
\bauthor{\bsnm{{Davies}}, \binits{R.I.}},
\bauthor{\bsnm{{Conan}}, \binits{J.-M.}}:
\bjtitle{\aap}
\bvolume{531},
\bfpage{151}
(\byear{2011}).
\arxivurl{1105.3455}.
doi:\doiurl{10.1051/0004-6361/201116603}
\end{barticle}
\endbibitem

\bibitem[\protect\citeauthoryear{{Fiorentino} et~al.}{2002}]{fiorentino02}
\begin{barticle}
\bauthor{\bsnm{{Fiorentino}}, \binits{G.}},
\bauthor{\bsnm{{Caputo}}, \binits{F.}},
\bauthor{\bsnm{{Marconi}}, \binits{M.}},
\bauthor{\bsnm{{Musella}}, \binits{I.}}:
\bjtitle{\apj}
\bvolume{576},
\bfpage{402}
(\byear{2002}).
\arxivurl{arXiv:astro-ph/0205147}.
doi:\doiurl{10.1086/341731}
\end{barticle}
\endbibitem

\bibitem[\protect\citeauthoryear{{Fiorentino} et~al.}{2007}]{fiorentino07}
\begin{barticle}
\bauthor{\bsnm{{Fiorentino}}, \binits{G.}},
\bauthor{\bsnm{{Marconi}}, \binits{M.}},
\bauthor{\bsnm{{Musella}}, \binits{I.}},
\bauthor{\bsnm{{Caputo}}, \binits{F.}}:
\bjtitle{\aap}
\bvolume{476},
\bfpage{863}
(\byear{2007}).
\arxivurl{0707.0959}.
doi:\doiurl{10.1051/0004-6361:20077587}
\end{barticle}
\endbibitem

\bibitem[\protect\citeauthoryear{{Fiorentino} et~al.}{2010}]{fiorentino10a}
\begin{barticle}
\bauthor{\bsnm{{Fiorentino}}, \binits{G.}},
\bauthor{\bsnm{{Contreras Ramos}}, \binits{R.}},
\bauthor{\bsnm{{Clementini}}, \binits{G.}},
\bauthor{\bsnm{{Marconi}}, \binits{M.}},
\bauthor{\bsnm{{Musella}}, \binits{I.}},
\bauthor{\bsnm{{Aloisi}}, \binits{A.}},
\bauthor{\bsnm{{Annibali}}, \binits{F.}},
\bauthor{\bsnm{{Saha}}, \binits{A.}},
\bauthor{\bsnm{{Tosi}}, \binits{M.}},
\bauthor{\bsnm{{van der Marel}}, \binits{R.P.}}:
\bjtitle{\apj}
\bvolume{711},
\bfpage{808}
(\byear{2010}).
\arxivurl{1001.4044}.
doi:\doiurl{10.1088/0004-637X/711/2/808}
\end{barticle}
\endbibitem

\bibitem[\protect\citeauthoryear{{Fiorentino} et~al.}{2011}]{fiorentino11}
\begin{barticle}
\bauthor{\bsnm{{Fiorentino}}, \binits{G.}},
\bauthor{\bsnm{{Tolstoy}}, \binits{E.}},
\bauthor{\bsnm{{Diolaiti}}, \binits{E.}},
\bauthor{\bsnm{{Valenti}}, \binits{E.}},
\bauthor{\bsnm{{Cignoni}}, \binits{M.}},
\bauthor{\bsnm{{Mackey}}, \binits{A.D.}}:
\bjtitle{\aap}
\bvolume{535},
\bfpage{63}
(\byear{2011}).
\arxivurl{1109.0161}.
doi:\doiurl{10.1051/0004-6361/201016094}
\end{barticle}
\endbibitem

\bibitem[\protect\citeauthoryear{{Freedman} et~al.}{1985}]{freedman85}
\begin{barticle}
\bauthor{\bsnm{{Freedman}}, \binits{W.L.}},
\bauthor{\bsnm{{Grieve}}, \binits{G.R.}},
\bauthor{\bsnm{{Madore}}, \binits{B.F.}}:
\bjtitle{\apjs}
\bvolume{59},
\bfpage{311}
(\byear{1985}).
doi:\doiurl{10.1086/191074}
\end{barticle}
\endbibitem

\bibitem[\protect\citeauthoryear{{Freedman} et~al.}{2001}]{freedman01}
\begin{barticle}
\bauthor{\bsnm{{Freedman}}, \binits{W.L.}},
\bauthor{\bsnm{{Madore}}, \binits{B.F.}},
\bauthor{\bsnm{{Gibson}}, \binits{B.K.}},
\bauthor{\bsnm{{Ferrarese}}, \binits{L.}},
\bauthor{\bsnm{{Kelson}}, \binits{D.D.}},
\bauthor{\bsnm{{Sakai}}, \binits{S.}},
\bauthor{\bsnm{{Mould}}, \binits{J.R.}},
\bauthor{\bsnm{{Kennicutt}}, \binits{R.C.} \bsuffix{Jr.}},
\bauthor{\bsnm{{Ford}}, \binits{H.C.}},
\bauthor{\bsnm{{Graham}}, \binits{J.A.}},
\bauthor{\bsnm{{Huchra}}, \binits{J.P.}},
\bauthor{\bsnm{{Hughes}}, \binits{S.M.G.}},
\bauthor{\bsnm{{Illingworth}}, \binits{G.D.}},
\bauthor{\bsnm{{Macri}}, \binits{L.M.}},
\bauthor{\bsnm{{Stetson}}, \binits{P.B.}}:
\bjtitle{\apj}
\bvolume{553},
\bfpage{47}
(\byear{2001}).
\arxivurl{arXiv:astro-ph/0012376}.
doi:\doiurl{10.1086/320638}
\end{barticle}
\endbibitem

\bibitem[\protect\citeauthoryear{{Gerke} et~al.}{2011}]{gerke11}
\begin{barticle}
\bauthor{\bsnm{{Gerke}}, \binits{J.R.}},
\bauthor{\bsnm{{Kochanek}}, \binits{C.S.}},
\bauthor{\bsnm{{Prieto}}, \binits{J.L.}},
\bauthor{\bsnm{{Stanek}}, \binits{K.Z.}},
\bauthor{\bsnm{{Macri}}, \binits{L.M.}}:
\bjtitle{\apj}
\bvolume{743},
\bfpage{176}
(\byear{2011}).
\arxivurl{1103.0549}.
doi:\doiurl{10.1088/0004-637X/743/2/176}
\end{barticle}
\endbibitem

\bibitem[\protect\citeauthoryear{{Gieren} et~al.}{2004}]{gieren04}
\begin{barticle}
\bauthor{\bsnm{{Gieren}}, \binits{W.}},
\bauthor{\bsnm{{Pietrzy{\'n}ski}}, \binits{G.}},
\bauthor{\bsnm{{Walker}}, \binits{A.}},
\bauthor{\bsnm{{Bresolin}}, \binits{F.}},
\bauthor{\bsnm{{Minniti}}, \binits{D.}},
\bauthor{\bsnm{{Kudritzki}}, \binits{R.-P.}},
\bauthor{\bsnm{{Udalski}}, \binits{A.}},
\bauthor{\bsnm{{Soszy{\'n}ski}}, \binits{I.}},
\bauthor{\bsnm{{Fouqu{\'e}}}, \binits{P.}},
\bauthor{\bsnm{{Storm}}, \binits{J.}},
\bauthor{\bsnm{{Bono}}, \binits{G.}}:
\bjtitle{\aj}
\bvolume{128},
\bfpage{1167}
(\byear{2004}).
\arxivurl{arXiv:astro-ph/0405581}.
doi:\doiurl{10.1086/422924}
\end{barticle}
\endbibitem

\bibitem[\protect\citeauthoryear{{Gieren} et~al.}{2005}]{gieren05}
\begin{barticle}
\bauthor{\bsnm{{Gieren}}, \binits{W.}},
\bauthor{\bsnm{{Pietrzy{\'n}ski}}, \binits{G.}},
\bauthor{\bsnm{{Soszy{\'n}ski}}, \binits{I.}},
\bauthor{\bsnm{{Bresolin}}, \binits{F.}},
\bauthor{\bsnm{{Kudritzki}}, \binits{R.-P.}},
\bauthor{\bsnm{{Minniti}}, \binits{D.}},
\bauthor{\bsnm{{Storm}}, \binits{J.}}:
\bjtitle{\apj}
\bvolume{628},
\bfpage{695}
(\byear{2005}).
\arxivurl{arXiv:astro-ph/0503626}.
doi:\doiurl{10.1086/430903}
\end{barticle}
\endbibitem

\bibitem[\protect\citeauthoryear{{Hilditch} et~al.}{2005}]{hilditch05}
\begin{barticle}
\bauthor{\bsnm{{Hilditch}}, \binits{R.W.}},
\bauthor{\bsnm{{Howarth}}, \binits{I.D.}},
\bauthor{\bsnm{{Harries}}, \binits{T.J.}}:
\bjtitle{\mnras}
\bvolume{357},
\bfpage{304}
(\byear{2005}).
\arxivurl{arXiv:astro-ph/0411672}.
doi:\doiurl{10.1111/j.1365-2966.2005.08653.x}
\end{barticle}
\endbibitem

\bibitem[\protect\citeauthoryear{{Hoessel} et~al.}{1998}]{hoessel98}
\begin{barticle}
\bauthor{\bsnm{{Hoessel}}, \binits{J.G.}},
\bauthor{\bsnm{{Saha}}, \binits{A.}},
\bauthor{\bsnm{{Danielson}}, \binits{G.E.}}:
\bjtitle{\aj}
\bvolume{116},
\bfpage{1679}
(\byear{1998}).
doi:\doiurl{10.1086/300528}
\end{barticle}
\endbibitem

\bibitem[\protect\citeauthoryear{{Keller} and {Wood}}{2006}]{keller06}
\begin{barticle}
\bauthor{\bsnm{{Keller}}, \binits{S.C.}},
\bauthor{\bsnm{{Wood}}, \binits{P.R.}}:
\bjtitle{\apj}
\bvolume{642},
\bfpage{834}
(\byear{2006}).
\arxivurl{arXiv:astro-ph/0601225}.
doi:\doiurl{10.1086/501115}
\end{barticle}
\endbibitem

\bibitem[\protect\citeauthoryear{{Kennicutt} et~al.}{1998}]{kennicutt98}
\begin{barticle}
\bauthor{\bsnm{{Kennicutt}}, \binits{R.C.} \bsuffix{Jr.}},
\bauthor{\bsnm{{Stetson}}, \binits{P.B.}},
\bauthor{\bsnm{{Saha}}, \binits{A.}},
\bauthor{\bsnm{{Kelson}}, \binits{D.}},
\bauthor{\bsnm{{Rawson}}, \binits{D.M.}},
\bauthor{\bsnm{{Sakai}}, \binits{S.}},
\bauthor{\bsnm{{Madore}}, \binits{B.F.}},
\bauthor{\bsnm{{Mould}}, \binits{J.R.}},
\bauthor{\bsnm{{Freedman}}, \binits{W.L.}},
\bauthor{\bsnm{{Bresolin}}, \binits{F.}},
\bauthor{\bsnm{{Ferrarese}}, \binits{L.}},
\bauthor{\bsnm{{Ford}}, \binits{H.}},
\bauthor{\bsnm{{Gibson}}, \binits{B.K.}},
\bauthor{\bsnm{{Graham}}, \binits{J.A.}},
\bauthor{\bsnm{{Han}}, \binits{M.}},
\bauthor{\bsnm{{Harding}}, \binits{P.}},
\bauthor{\bsnm{{Hoessel}}, \binits{J.G.}},
\bauthor{\bsnm{{Huchra}}, \binits{J.P.}},
\bauthor{\bsnm{{Hughes}}, \binits{S.M.G.}},
\bauthor{\bsnm{{Illingworth}}, \binits{G.D.}},
\bauthor{\bsnm{{Macri}}, \binits{L.M.}},
\bauthor{\bsnm{{Phelps}}, \binits{R.L.}},
\bauthor{\bsnm{{Silbermann}}, \binits{N.A.}},
\bauthor{\bsnm{{Turner}}, \binits{A.M.}},
\bauthor{\bsnm{{Wood}}, \binits{P.R.}}:
\bjtitle{\apj}
\bvolume{498},
\bfpage{181}
(\byear{1998}).
\arxivurl{arXiv:astro-ph/9712055}.
doi:\doiurl{10.1086/305538}
\end{barticle}
\endbibitem

\bibitem[\protect\citeauthoryear{{Komatsu} et~al.}{2011}]{komatsu11}
\begin{barticle}
\bauthor{\bsnm{{Komatsu}}, \binits{E.}},
\bauthor{\bsnm{{Smith}}, \binits{K.M.}},
\bauthor{\bsnm{{Dunkley}}, \binits{J.}},
\bauthor{\bsnm{{Bennett}}, \binits{C.L.}},
\bauthor{\bsnm{{Gold}}, \binits{B.}},
\bauthor{\bsnm{{Hinshaw}}, \binits{G.}},
\bauthor{\bsnm{{Jarosik}}, \binits{N.}},
\bauthor{\bsnm{{Larson}}, \binits{D.}},
\bauthor{\bsnm{{Nolta}}, \binits{M.R.}},
\bauthor{\bsnm{{Page}}, \binits{L.}},
\bauthor{\bsnm{{Spergel}}, \binits{D.N.}},
\bauthor{\bsnm{{Halpern}}, \binits{M.}},
\bauthor{\bsnm{{Hill}}, \binits{R.S.}},
\bauthor{\bsnm{{Kogut}}, \binits{A.}},
\bauthor{\bsnm{{Limon}}, \binits{M.}},
\bauthor{\bsnm{{Meyer}}, \binits{S.S.}},
\bauthor{\bsnm{{Odegard}}, \binits{N.}},
\bauthor{\bsnm{{Tucker}}, \binits{G.S.}},
\bauthor{\bsnm{{Weiland}}, \binits{J.L.}},
\bauthor{\bsnm{{Wollack}}, \binits{E.}},
\bauthor{\bsnm{{Wright}}, \binits{E.L.}}:
\bjtitle{\apjs}
\bvolume{192},
\bfpage{18}
(\byear{2011}).
\arxivurl{1001.4538}.
doi:\doiurl{10.1088/0067-0049/192/2/18}
\end{barticle}
\endbibitem

\bibitem[\protect\citeauthoryear{{Madore}}{1976}]{madore76}
\begin{bchapter}
\bauthor{\bsnm{{Madore}}, \binits{B.F.}}:
In: \beditor{\bsnm{{R.~J.~Dickens, J.~E.~Perry, F.~G.~Smith, \& I.~R.~King}}}
  (ed.)
\bbtitle{The Galaxy and the Local Group}.
\bsertitle{Royal Greenwich Observatory Bulletins},
vol. \bseriesno{182},
p. \bfpage{153}
(\byear{1976})
\end{bchapter}
\endbibitem

\bibitem[\protect\citeauthoryear{{Marconi} et~al.}{2005}]{marconi05}
\begin{barticle}
\bauthor{\bsnm{{Marconi}}, \binits{M.}},
\bauthor{\bsnm{{Musella}}, \binits{I.}},
\bauthor{\bsnm{{Fiorentino}}, \binits{G.}}:
\bjtitle{\apj}
\bvolume{632},
\bfpage{590}
(\byear{2005}).
\arxivurl{arXiv:astro-ph/0506207}.
doi:\doiurl{10.1086/432790}
\end{barticle}
\endbibitem

\bibitem[\protect\citeauthoryear{{Marconi} et~al.}{2010}]{marconi10}
\begin{barticle}
\bauthor{\bsnm{{Marconi}}, \binits{M.}},
\bauthor{\bsnm{{Musella}}, \binits{I.}},
\bauthor{\bsnm{{Fiorentino}}, \binits{G.}},
\bauthor{\bsnm{{Clementini}}, \binits{G.}},
\bauthor{\bsnm{{Aloisi}}, \binits{A.}},
\bauthor{\bsnm{{Annibali}}, \binits{F.}},
\bauthor{\bsnm{{Contreras Ramos}}, \binits{R.}},
\bauthor{\bsnm{{Saha}}, \binits{A.}},
\bauthor{\bsnm{{Tosi}}, \binits{M.}},
\bauthor{\bsnm{{van der Marel}}, \binits{R.P.}}:
\bjtitle{\apj}
\bvolume{713},
\bfpage{615}
(\byear{2010}).
\arxivurl{1002.4752}.
doi:\doiurl{10.1088/0004-637X/713/1/615}
\end{barticle}
\endbibitem

\bibitem[\protect\citeauthoryear{{Pagel} et~al.}{1978}]{pagel78}
\begin{barticle}
\bauthor{\bsnm{{Pagel}}, \binits{B.E.J.}},
\bauthor{\bsnm{{Edmunds}}, \binits{M.G.}},
\bauthor{\bsnm{{Fosbury}}, \binits{R.A.E.}},
\bauthor{\bsnm{{Webster}}, \binits{B.L.}}:
\bjtitle{\mnras}
\bvolume{184},
\bfpage{569}
(\byear{1978})
\end{barticle}
\endbibitem

\bibitem[\protect\citeauthoryear{{Pedicelli} et~al.}{2009}]{pedicelli09}
\begin{barticle}
\bauthor{\bsnm{{Pedicelli}}, \binits{S.}},
\bauthor{\bsnm{{Bono}}, \binits{G.}},
\bauthor{\bsnm{{Lemasle}}, \binits{B.}},
\bauthor{\bsnm{{Fran{\c c}ois}}, \binits{P.}},
\bauthor{\bsnm{{Groenewegen}}, \binits{M.}},
\bauthor{\bsnm{{Lub}}, \binits{J.}},
\bauthor{\bsnm{{Pel}}, \binits{J.W.}},
\bauthor{\bsnm{{Laney}}, \binits{D.}},
\bauthor{\bsnm{{Piersimoni}}, \binits{A.}},
\bauthor{\bsnm{{Romaniello}}, \binits{M.}},
\bauthor{\bsnm{{Buonanno}}, \binits{R.}},
\bauthor{\bsnm{{Caputo}}, \binits{F.}},
\bauthor{\bsnm{{Cassisi}}, \binits{S.}},
\bauthor{\bsnm{{Castelli}}, \binits{F.}},
\bauthor{\bsnm{{Leurini}}, \binits{S.}},
\bauthor{\bsnm{{Pietrinferni}}, \binits{A.}},
\bauthor{\bsnm{{Primas}}, \binits{F.}},
\bauthor{\bsnm{{Pritchard}}, \binits{J.}}:
\bjtitle{\aap}
\bvolume{504},
\bfpage{81}
(\byear{2009}).
\arxivurl{0906.3140}.
doi:\doiurl{10.1051/0004-6361/200912504}
\end{barticle}
\endbibitem

\bibitem[\protect\citeauthoryear{{Peimbert} and
  {Torres-Peimbert}}{1976}]{peimbert76}
\begin{barticle}
\bauthor{\bsnm{{Peimbert}}, \binits{M.}},
\bauthor{\bsnm{{Torres-Peimbert}}, \binits{S.}}:
\bjtitle{\apj}
\bvolume{203},
\bfpage{581}
(\byear{1976}).
doi:\doiurl{10.1086/154114}
\end{barticle}
\endbibitem

\bibitem[\protect\citeauthoryear{{Pietrzy{\'n}ski}
  et~al.}{2004}]{pietrzynski04}
\begin{barticle}
\bauthor{\bsnm{{Pietrzy{\'n}ski}}, \binits{G.}},
\bauthor{\bsnm{{Gieren}}, \binits{W.}},
\bauthor{\bsnm{{Udalski}}, \binits{A.}},
\bauthor{\bsnm{{Bresolin}}, \binits{F.}},
\bauthor{\bsnm{{Kudritzki}}, \binits{R.-P.}},
\bauthor{\bsnm{{Soszy{\'n}ski}}, \binits{I.}},
\bauthor{\bsnm{{Szyma{\'n}ski}}, \binits{M.}},
\bauthor{\bsnm{{Kubiak}}, \binits{M.}}:
\bjtitle{\aj}
\bvolume{128},
\bfpage{2815}
(\byear{2004}).
\arxivurl{arXiv:astro-ph/0408572}.
doi:\doiurl{10.1086/425531}
\end{barticle}
\endbibitem

\bibitem[\protect\citeauthoryear{{Pietrzy{\'n}ski}
  et~al.}{2006}]{pietrzynski06}
\begin{barticle}
\bauthor{\bsnm{{Pietrzy{\'n}ski}}, \binits{G.}},
\bauthor{\bsnm{{Gieren}}, \binits{W.}},
\bauthor{\bsnm{{Soszy{\'n}ski}}, \binits{I.}},
\bauthor{\bsnm{{Udalski}}, \binits{A.}},
\bauthor{\bsnm{{Bresolin}}, \binits{F.}},
\bauthor{\bsnm{{Kudritzki}}, \binits{R.-P.}},
\bauthor{\bsnm{{Mennickent}}, \binits{R.}},
\bauthor{\bsnm{{Walker}}, \binits{A.}},
\bauthor{\bsnm{{Garcia}}, \binits{A.}},
\bauthor{\bsnm{{Szewczyk}}, \binits{O.}},
\bauthor{\bsnm{{Szyma{\'n}ski}}, \binits{M.}},
\bauthor{\bsnm{{Kubiak}}, \binits{M.}},
\bauthor{\bsnm{{Wyrzykowski}}, \binits{{\L}.}}:
\bjtitle{\aj}
\bvolume{132},
\bfpage{2556}
(\byear{2006}).
\arxivurl{arXiv:astro-ph/0610595}.
doi:\doiurl{10.1086/508927}
\end{barticle}
\endbibitem

\bibitem[\protect\citeauthoryear{{Riess} et~al.}{2009}]{riess09}
\begin{barticle}
\bauthor{\bsnm{{Riess}}, \binits{A.G.}},
\bauthor{\bsnm{{Macri}}, \binits{L.}},
\bauthor{\bsnm{{Li}}, \binits{W.}},
\bauthor{\bsnm{{Lampeitl}}, \binits{H.}},
\bauthor{\bsnm{{Casertano}}, \binits{S.}},
\bauthor{\bsnm{{Ferguson}}, \binits{H.C.}},
\bauthor{\bsnm{{Filippenko}}, \binits{A.V.}},
\bauthor{\bsnm{{Jha}}, \binits{S.W.}},
\bauthor{\bsnm{{Chornock}}, \binits{R.}},
\bauthor{\bsnm{{Greenhill}}, \binits{L.}},
\bauthor{\bsnm{{Mutchler}}, \binits{M.}},
\bauthor{\bsnm{{Ganeshalingham}}, \binits{M.}},
\bauthor{\bsnm{{Hicken}}, \binits{M.}}:
\bjtitle{\apjs}
\bvolume{183},
\bfpage{109}
(\byear{2009}).
\arxivurl{0905.0697}.
doi:\doiurl{10.1088/0067-0049/183/1/109}
\end{barticle}
\endbibitem

\bibitem[\protect\citeauthoryear{{Riess} et~al.}{2011}]{riess11a}
\begin{barticle}
\bauthor{\bsnm{{Riess}}, \binits{A.G.}},
\bauthor{\bsnm{{Macri}}, \binits{L.}},
\bauthor{\bsnm{{Casertano}}, \binits{S.}},
\bauthor{\bsnm{{Lampeitl}}, \binits{H.}},
\bauthor{\bsnm{{Ferguson}}, \binits{H.C.}},
\bauthor{\bsnm{{Filippenko}}, \binits{A.V.}},
\bauthor{\bsnm{{Jha}}, \binits{S.W.}},
\bauthor{\bsnm{{Li}}, \binits{W.}},
\bauthor{\bsnm{{Chornock}}, \binits{R.}}:
\bjtitle{\apj}
\bvolume{730},
\bfpage{119}
(\byear{2011}).
\arxivurl{1103.2976}.
doi:\doiurl{10.1088/0004-637X/730/2/119}
\end{barticle}
\endbibitem

\bibitem[\protect\citeauthoryear{{Romaniello} et~al.}{2008}]{romaniello08}
\begin{barticle}
\bauthor{\bsnm{{Romaniello}}, \binits{M.}},
\bauthor{\bsnm{{Primas}}, \binits{F.}},
\bauthor{\bsnm{{Mottini}}, \binits{M.}},
\bauthor{\bsnm{{Pedicelli}}, \binits{S.}},
\bauthor{\bsnm{{Lemasle}}, \binits{B.}},
\bauthor{\bsnm{{Bono}}, \binits{G.}},
\bauthor{\bsnm{{Fran{\c c}ois}}, \binits{P.}},
\bauthor{\bsnm{{Groenewegen}}, \binits{M.A.T.}},
\bauthor{\bsnm{{Laney}}, \binits{C.D.}}:
\bjtitle{\aap}
\bvolume{488},
\bfpage{731}
(\byear{2008}).
\arxivurl{0807.1196}.
doi:\doiurl{10.1051/0004-6361:20065661}
\end{barticle}
\endbibitem

\bibitem[\protect\citeauthoryear{{Saha} et~al.}{2001}]{saha01}
\begin{barticle}
\bauthor{\bsnm{{Saha}}, \binits{A.}},
\bauthor{\bsnm{{Sandage}}, \binits{A.}},
\bauthor{\bsnm{{Thim}}, \binits{F.}},
\bauthor{\bsnm{{Labhardt}}, \binits{L.}},
\bauthor{\bsnm{{Tammann}}, \binits{G.A.}},
\bauthor{\bsnm{{Christensen}}, \binits{J.}},
\bauthor{\bsnm{{Panagia}}, \binits{N.}},
\bauthor{\bsnm{{Macchetto}}, \binits{F.D.}}:
\bjtitle{\apj}
\bvolume{551},
\bfpage{973}
(\byear{2001}).
\arxivurl{arXiv:astro-ph/0012015}.
doi:\doiurl{10.1086/320223}
\end{barticle}
\endbibitem

\bibitem[\protect\citeauthoryear{{Sakai} et~al.}{2004}]{sakai04}
\begin{barticle}
\bauthor{\bsnm{{Sakai}}, \binits{S.}},
\bauthor{\bsnm{{Ferrarese}}, \binits{L.}},
\bauthor{\bsnm{{Kennicutt}}, \binits{R.C.} \bsuffix{Jr.}},
\bauthor{\bsnm{{Saha}}, \binits{A.}}:
\bjtitle{\apj}
\bvolume{608},
\bfpage{42}
(\byear{2004}).
\arxivurl{arXiv:astro-ph/0402499}.
doi:\doiurl{10.1086/386540}
\end{barticle}
\endbibitem

\bibitem[\protect\citeauthoryear{{Sandage} et~al.}{2009}]{sandage09}
\begin{barticle}
\bauthor{\bsnm{{Sandage}}, \binits{A.}},
\bauthor{\bsnm{{Tammann}}, \binits{G.A.}},
\bauthor{\bsnm{{Reindl}}, \binits{B.}}:
\bjtitle{\aap}
\bvolume{493},
\bfpage{471}
(\byear{2009}).
\arxivurl{0810.1780}.
doi:\doiurl{10.1051/0004-6361:200810550}
\end{barticle}
\endbibitem

\bibitem[\protect\citeauthoryear{{Soszy{\~n}ski} et~al.}{2010}]{soszynski10a}
\begin{barticle}
\bauthor{\bsnm{{Soszy{\~n}ski}}, \binits{I.}},
\bauthor{\bsnm{{Poleski}}, \binits{R.}},
\bauthor{\bsnm{{Udalski}}, \binits{A.}},
\bauthor{\bsnm{{Szyma{\~n}ski}}, \binits{M.K.}},
\bauthor{\bsnm{{Kubiak}}, \binits{M.}},
\bauthor{\bsnm{{Pietrzy{\~n}ski}}, \binits{G.}},
\bauthor{\bsnm{{Wyrzykowski}}, \binits{{\L}.}},
\bauthor{\bsnm{{Szewczyk}}, \binits{O.}},
\bauthor{\bsnm{{Ulaczyk}}, \binits{K.}}:
\bjtitle{\actaa}
\bvolume{60},
\bfpage{17}
(\byear{2010}).
\arxivurl{1003.4518}
\end{barticle}
\endbibitem

\bibitem[\protect\citeauthoryear{{Soszy{\'n}ski} et~al.}{2008}]{soszynski08b}
\begin{barticle}
\bauthor{\bsnm{{Soszy{\'n}ski}}, \binits{I.}},
\bauthor{\bsnm{{Poleski}}, \binits{R.}},
\bauthor{\bsnm{{Udalski}}, \binits{A.}},
\bauthor{\bsnm{{Szymanski}}, \binits{M.K.}},
\bauthor{\bsnm{{Kubiak}}, \binits{M.}},
\bauthor{\bsnm{{Pietrzynski}}, \binits{G.}},
\bauthor{\bsnm{{Wyrzykowski}}, \binits{L.}},
\bauthor{\bsnm{{Szewczyk}}, \binits{O.}},
\bauthor{\bsnm{{Ulaczyk}}, \binits{K.}}:
\bjtitle{\actaa}
\bvolume{58},
\bfpage{163}
(\byear{2008}).
\arxivurl{0808.2210}
\end{barticle}
\endbibitem

\bibitem[\protect\citeauthoryear{{Storm} et~al.}{2011}]{storm11}
\begin{barticle}
\bauthor{\bsnm{{Storm}}, \binits{J.}},
\bauthor{\bsnm{{Gieren}}, \binits{W.}},
\bauthor{\bsnm{{Fouqu{\'e}}}, \binits{P.}},
\bauthor{\bsnm{{Barnes}}, \binits{T.G.}},
\bauthor{\bsnm{{Soszy{\'n}ski}}, \binits{I.}},
\bauthor{\bsnm{{Pietrzy{\'n}ski}}, \binits{G.}},
\bauthor{\bsnm{{Nardetto}}, \binits{N.}},
\bauthor{\bsnm{{Queloz}}, \binits{D.}}:
\bjtitle{\aap}
\bvolume{534},
\bfpage{95}
(\byear{2011}).
\arxivurl{1109.2016}.
doi:\doiurl{10.1051/0004-6361/201117154}
\end{barticle}
\endbibitem

\bibitem[\protect\citeauthoryear{{Tolstoy} et~al.}{2010}]{tolstoy10}
\begin{botherref}
\oauthor{\bsnm{{Tolstoy}}, \binits{E.}},
\oauthor{\bsnm{{Battaglia}}, \binits{G.}},
\oauthor{\bsnm{{Beck}}, \binits{R.}},
\oauthor{\bsnm{{Brunthaler}}, \binits{A.}},
\oauthor{\bsnm{{Calamida}}, \binits{A.}},
\oauthor{\bsnm{{Fiorentino}}, \binits{G.}},
\oauthor{\bsnm{{van der Hulst}}, \binits{J.M.}}:
ArXiv e-prints
(2010).
\arxivurl{1009.4103}
\end{botherref}
\endbibitem

\bibitem[\protect\citeauthoryear{{Tosi} et~al.}{2001}]{tosi01}
\begin{barticle}
\bauthor{\bsnm{{Tosi}}, \binits{M.}},
\bauthor{\bsnm{{Sabbi}}, \binits{E.}},
\bauthor{\bsnm{{Bellazzini}}, \binits{M.}},
\bauthor{\bsnm{{Aloisi}}, \binits{A.}},
\bauthor{\bsnm{{Greggio}}, \binits{L.}},
\bauthor{\bsnm{{Leitherer}}, \binits{C.}},
\bauthor{\bsnm{{Montegriffo}}, \binits{P.}}:
\bjtitle{\aj}
\bvolume{122},
\bfpage{1271}
(\byear{2001}).
\arxivurl{arXiv:astro-ph/0105357}.
doi:\doiurl{10.1086/322104}
\end{barticle}
\endbibitem

\bibitem[\protect\citeauthoryear{{Udalski} et~al.}{1999}]{udalski99}
\begin{barticle}
\bauthor{\bsnm{{Udalski}}, \binits{A.}},
\bauthor{\bsnm{{Soszynski}}, \binits{I.}},
\bauthor{\bsnm{{Szymanski}}, \binits{M.}},
\bauthor{\bsnm{{Kubiak}}, \binits{M.}},
\bauthor{\bsnm{{Pietrzynski}}, \binits{G.}},
\bauthor{\bsnm{{Wozniak}}, \binits{P.}},
\bauthor{\bsnm{{Zebrun}}, \binits{K.}}:
\bjtitle{\actaa}
\bvolume{49},
\bfpage{223}
(\byear{1999}).
\arxivurl{arXiv:astro-ph/9908317}
\end{barticle}
\endbibitem

\bibitem[\protect\citeauthoryear{{Urbaneja} et~al.}{2005}]{urbaneja05}
\begin{barticle}
\bauthor{\bsnm{{Urbaneja}}, \binits{M.A.}},
\bauthor{\bsnm{{Herrero}}, \binits{A.}},
\bauthor{\bsnm{{Bresolin}}, \binits{F.}},
\bauthor{\bsnm{{Kudritzki}}, \binits{R.-P.}},
\bauthor{\bsnm{{Gieren}}, \binits{W.}},
\bauthor{\bsnm{{Puls}}, \binits{J.}},
\bauthor{\bsnm{{Przybilla}}, \binits{N.}},
\bauthor{\bsnm{{Najarro}}, \binits{F.}},
\bauthor{\bsnm{{Pietrzy{\'n}ski}}, \binits{G.}}:
\bjtitle{\apj}
\bvolume{622},
\bfpage{862}
(\byear{2005}).
doi:\doiurl{10.1086/427468}
\end{barticle}
\endbibitem

\end{thebibliography}

\end{document}